\documentclass[english,showpacs,preprintnumbers,amsmath,amssymb]{revtex4}

\usepackage{amsfonts}
\usepackage[T1]{fontenc}
\usepackage[latin9]{inputenc}
\usepackage{graphicx}
\usepackage{amsmath}
\usepackage{amssymb}
\usepackage{esint}
\usepackage{bm}
\usepackage{babel}


\begin{document}

\title{Coulomb Screening of 2D Massive Dirac Fermions}

\author{Jia-Ning Zhang}\email{jnzhang05@gmail.com}

\affiliation{Chern Institute of Mathematics, Nankai University,
Tianjin 300071, China}

\begin{abstract}

A model of 2D massive Dirac fermions ,interacting with a
instantaneous $1/r$ Coulomb interaction, is presented to mimic the
physics of gapped graphene. The static polarization function is
calculated explicitly to analyze screening effect at the finite
temperature and density. Results are compared with the massless case
. We also show that various other works can be reproduced within our
model in a straightforward and unified manner.
\end{abstract}
 \maketitle \setlength{\baselineskip}{18pt}

\section{Introduction}

 \indent Over the past several years, the physics of graphene has
attracted considerable interest, both theoretically and
experimentally\cite{Geim2009RMP}. Graphene, due to its two
dimensional hexagonal lattice structure, has unique linear energy
spectrum near the Dirac points of Brillouin zone. Because of the
unusual energy band dispersion, many electronic properties in
graphene exhibit significantly different behavior from the
conventional 2D systems,
for example, half-integer QHE(Quantum Hall Effect)\cite{Geim2009NP}.\\
 \indent In graphene although the
motion of electrons are fixed on the 2D plane, their interactions
still show 3D Coulomb's law, for the electric field lines cannot be
confined on 2D. While most of the early work were based on massless
Dirac fermion model, recent work has shown that
 the massive case can also be created\cite{Geim2009NP}.\\
 \indent For massive fermions, it is equivalent to
 the opening of a gap in the electronic spectrum in condensed matter physics. So the similarity
 between graphene and $QED_{2+1}$ is obvious, and many results were
 obtained by exploring this correspondence\cite{Son2007PRB}. While the main difference is
 that Graphene lacks of Lorenz invariance due to its nearly
 instantaneous coulomb interaction.

 \indent Besides, it should be noted there has been many papers \cite{wunsch2006NJP}\cite{Hwang2007PRB}\cite{Vafek2007PRL}\cite{Asgari2007JPA}
 on the Coulomb interaction in gapless graphene. Ref.\cite{wunsch2006NJP}\cite{Hwang2007PRB} gave the
 polarization function for zero temperature gapless graphene at
 finite density. Ref.\cite{Vafek2007PRL} dealt with
 2D Coulomb-interacting massless Dirac fermions and calculated the specific
 heat at finite temperature, while Ref.\cite{Asgari2007JPA}
 generalized it to the finite density case. As for the massive Dirac
 fermions, there are also many work on it.
 However, most of them (gapless or gapped graphene) were highly succinct, used different models and did not give the calculations of polarization function in detail.

 In this paper, we consider a model of two-component Dirac fermions interacting through a three
dimensional instantaneous Coulomb interaction, and calculate the
polarization functions at finite
 temperature and finite density using finite temperature field techniques\cite{Zee2003QFT}.
 In fact quantum field theory at finite temperature or density is usually applied to study cosmology and astrophysics
  But its 2D spatial case has not been observed in nature. So the graphene provides a wonderful platform for establishing
  $QED_{2+1}$.  We also show various works on this topic can be connected within our
model in a straightforward and unified manner.

\section{Model}
 \indent Our starting point is a model of $2+1$ dimensional two-component
Dirac fermions mediated by a three dimensional Coulomb interaction
at temperature $T$. The action of the system $S$ is given by
$(\hbar=1)$
\begin{eqnarray}
S(\bar{\psi},\psi,\varphi)&=&\int_{0}^{\beta}d\tau\{\int d^{3}x
\frac{1}{8\pi}|\partial_{i}\varphi(x,\tau)|^{2}+\int
d^{2}x\sum_{s=1}^{N}\bar{\psi}_{s}(x,\tau)[\partial_{\tau}+v \bm{\sigma}\cdot \textbf{p}+m\sigma_{3}\nonumber\\
& &-\mu+ie\varphi(x,\tau)]\psi_{s}(x,\tau)\}\label{action}
\end{eqnarray}

 \indent Here, $\beta=1/T$ and $\mu$ is the chemical potential. The fields $\psi_{s}$ are two-component fermion fields, subscript
index $s$ stands for different species of fermions with $N=4$ due to
spin, valley degeneracies in graphene. The vector
$\bm{\sigma}=(\sigma_{1},\sigma_{2})$ and $\sigma_{i},i=1,2,3 $ are
pauli matrices; $\sigma_{0}=I$ is the $2\times2$ identity
matrix(omitted for simplicity). $\varphi$ is the field that mediates
the Coulomb interaction, nothing but the time component of the
electromagnetic field. The 3D space integral of the action describes
the kinetic term of the scalar field, while the remaining terms
describe kinetic term for the fermion fields and their interaction
with the scalar field. In addition, we put $v=1$ bellow and only
restore it if necessary.

Correspondingly, the Green's function for the free Dirac fermions
\begin{equation}
G_{0}(k)=\frac{k_{0}+\bm{\sigma}\cdot \textbf{k} +m
\sigma_{3}}{k_{0}^{2}-\textbf{k}^{2}-m^{2}}
\end{equation}

Using more symmetric three-momentum
notation,$(q_{0},\textbf{q})=(i\omega_{l},\textbf{q}),(k_{0},\textbf{k})=(i\omega_{n}+\mu,\textbf{k}),\tilde{k}^{2}=k_{0}^{2}-\textbf{k}^2,\omega_{n}=(2n+1)\pi/\beta,\omega_{l}=2l\pi/\beta$
etc.,then the polarization function in the random-phase
approximation (RPA)\cite{Mahan1993MPP} is

\begin{eqnarray}
\lefteqn{\Pi(q) =\frac{4}{\beta}\sum_{n}\int \frac{d^{2}\textbf{k}}{(2\pi)^{2}} Tr(G(\tilde{k})G(\tilde{k}+\tilde{q}))}\nonumber\\
& & =\frac{8}{\beta }\sum_{n}\int
\frac{d^{2}\textbf{k}}{(2\pi)^{2}}\frac{k_{0}(k_{0}+q_{0})+\textbf{k}(\textbf{k}+\textbf{q})+m^{2}}{(\tilde{k}^{2}-m^{2})[(\tilde{k}+\tilde{q})^{2}-m^{2}]}
\end{eqnarray}

where $v(\textbf{q})=2\pi e^{2}/\kappa|\textbf{q}|$ is the 2D
Fourier transform of the 3D Coulomb interaction. Following similar
consideration in Ref.\cite{Gale2006ftq} we can divide $\Pi$ into two
contributions, the vacuum part and matter part.

\begin{equation}
\Pi=\Pi_{vac}+\Pi_{matter}
\end{equation}

\begin{equation}
\lim_{T\rightarrow 0,\mu\rightarrow 0}\Pi=\Pi_{vac}
\end{equation}

 \indent The vacuum polarization function, both massless and massive case,
can be calculated by dimensional regularization approach as in
Ref.\cite{Son2007PRB}\cite{Kotov2008PRB}. Here, we can reproduce
their results within our model

\begin{equation}
\Pi_{vac}(q)=-\frac{|\textbf{q}|^{2}}{\pi}\{\frac{m}{q^{2}}+\frac{1}{2q}(1-\frac{4m^{2}}{q^{2}})\arctan(\frac{q}{2m})\}
\end{equation}

\section{Polarization function for massive Dirac fermions}
The sum of fermion Matsubara frequencies can be performed
in a standard manner\cite{Nagaosa1999QFTICMP}. \\

\begin{eqnarray}
&&\frac{1}{\beta}\sum_{n}\frac{k_{0}(k_{0}+q_{0})+\textbf{k}\cdot(\textbf{k}+\textbf{q})+m^{2}}{(k_{0}^{2}-E_{\textbf{k}}^{2})[(k_{0}+q_{0}^{2})-E_{\textbf{k}+\textbf{q}}^{2}]}\nonumber\\
&=&-\frac{1}{2\pi i}\oint dz h(z)g(z)\nonumber\\
&=&-\frac{1}{2\pi i}\int_{i\infty-\epsilon}^{-i\infty-\epsilon} dz
h(z) \frac{1}{2}\tanh(\beta z/2)-\frac{1}{2\pi
i}\int_{-i\infty+\epsilon}^{i\infty+\epsilon} dz h(z)
\frac{1}{2}\tanh(\beta z/2)\nonumber\\
&=&-\frac{1}{2\pi i}\int_{i\infty-\epsilon}^{-i\infty-\epsilon} dz
h(z) (\frac{1}{2}-\frac{1}{e^{-\beta z}+1})-\frac{1}{2\pi
i}\int_{-i\infty+\epsilon}^{i\infty+\epsilon} dz h(z)
(-\frac{1}{2}+\frac{1}{e^{\beta z}+1})
\end{eqnarray}
With
\begin{equation}
h(z)=\frac{(z+\mu)(z+\mu+q_{0})+\textbf{k}\cdot(\textbf{k}+\textbf{q})+m^{2}}{[(z+\mu)^{2}-E_{\textbf{k}}^{2}][(z+\mu+q_{0}^{2})-E_{\textbf{k}+\textbf{q}}^{2}]}\nonumber
\end{equation}
\begin{equation}
g(z)=\frac{\beta}{2}\tanh(\beta z/2)\nonumber
\end{equation}
\begin{equation}
E_{\textbf{k}}=\sqrt{\textbf{k}^{2}+m^{2}}\nonumber
\end{equation}
For function $g(z)$ has simple poles at $z=i\omega_{n}$, the sum
emerge as the integration of the product $hg$ along a suitable path
in the complex plane. We can divide the above expressions into two
parts $\pi_{vac},\pi_{matter}$, which describe the vacuum and
matter's contributions respectively.

\begin{equation} \pi_{vac}=\frac{1}{2\pi
i}\int_{-i\infty}^{i\infty} dz[h(z)+h(-z)]/2
\end{equation}

\begin{eqnarray}
&\pi_{matter}&=-\frac{1}{2\pi
i}\int_{-i\infty+\epsilon}^{i\infty+\epsilon} dz
h(z)\frac{1}{e^{\beta z}+1}+\frac{1}{2\pi
i}\int_{i\infty-\epsilon}^{-i\infty-\epsilon} dz
h(z)\frac{1}{e^{-\beta z}+1}\nonumber\\
&&=\frac{1}{2E_{\textbf{k}}}\frac{E_{\textbf{k}}(E_{\textbf{k}}+q_{0})+\textbf{k}(\textbf{k}+\textbf{q})+m^{2}}{(E_{\textbf{k}}+q_{0})^2-E_{\textbf{k}+\textbf{q}}^{2}}
\frac{1}{e^{\beta (E_{\textbf{k}}+\mu)}+1}\nonumber\\
&&+\frac{1}{2E_{\textbf{k}}}\frac{E_{\textbf{k}}(E_{\textbf{k}}-q_{0})+\textbf{k}(\textbf{k}+\textbf{q})+m^{2}}{(E_{\textbf{k}}-q_{0})^2-E_{\textbf{k}+\textbf{q}}^{2}}
\frac{1}{e^{\beta (E_{\textbf{k}}-\mu)}+1}\nonumber\\
&&{}+\frac{1}{2E_{\textbf{k}+\textbf{q}}}\frac{E_{\textbf{k}+\textbf{q}}(E_{\textbf{k}+\textbf{q}}+q_{0})+\textbf{k}(\textbf{k}+\textbf{q})+m^{2}}{(E_{\textbf{k}+\textbf{q}}+q_{0})^2-E_{\textbf{k}}^{2}}
\frac{1}{e^{\beta (E_{\textbf{k}+\textbf{q}}-\mu)}+1}\nonumber\\
&&{}+\frac{1}{2E_{\textbf{k}+\textbf{q}}}\frac{E_{\textbf{k}+\textbf{q}}(E_{\textbf{k}+\textbf{q}}-q_{0})+\textbf{k}(\textbf{k}+\textbf{q})+m^{2}}{(E_{\textbf{k}+\textbf{q}}-q_{0})^2-E_{\textbf{k}}^{2}}
\frac{1}{e^{\beta (E_{\textbf{k}+\textbf{q}}+\mu)}+1}\nonumber\\
&&=\frac{1}{2E_{\textbf{k}}}[f(q_{0})+f(-q_{0})]N_{F}(E_{\textbf{k}})
\end{eqnarray}
Where we have defined
\begin{equation}
f(q_{0})=\frac{E_{\textbf{k}}(E_{\textbf{k}}-q_{0})+\textbf{k}(\textbf{k}+\textbf{q})+m^{2}}{(E_{\textbf{k}}-q_{0})^2-E_{\textbf{k}+\textbf{q}}^{2}}\nonumber
\end{equation}
\begin{equation}
N_{F}(E_{\textbf{k}})=\frac{1}{e^{\beta(E_{\textbf{k}}+\mu)}+1}+\frac{1}{e^{\beta(E_{\textbf{k}}-\mu)}+1}\nonumber
\end{equation}

For vacuum part where we obtained in last section, it describes the
intrinsic graphene, in which the conduction band is empty while the
valence band is fully occupied at zero temperature. When
$m\rightarrow0$ graphene varies from a insulator to the zero-gap
semiconductor system. When we take into account the finite density
effect, the Fermi energy could lie either in valence band $(\mu<0)$
or in conduction band$(\mu>0)$

\begin{eqnarray}
&\Pi_{matter}&=8\int\frac{d^{2}\textbf{k}}{(2\pi)^{2}}\frac{1}{2E_{\textbf{k}}}[f(q_{0})+f(-q_{0})]\Theta(E_{\textbf{k}}-\mu)\nonumber\\
&&=\frac{1}{\pi^{2}}Re\int_{0}^{k_{F}}\frac{d k k}{E_{\textbf{k}}}
\int_{-1}^{1}dx\frac{1}{\sqrt{1-x^{2}}}(\frac{4E_{\textbf{k}}^{2}-4q_{0}E_{\textbf{k}}+q_{0}^{2}-q^{2}}
{q_{0}^{2}-2q_{0}E_{\textbf{k}}-q^{2}-2kq x}-1)
\end{eqnarray}
The retarded polarization function of the free fermions

\begin{equation}
\Im m\Pi_{matter}^{ret}=\frac{1}{\pi}Re\int_{0}^{k_{F}}\frac{d
k k}{E_{\textbf{k}}}\int_{-1}^{1} dx
\frac{4E_{\textbf{k}}^{2}-4q_{0}E_{\textbf{k}}+q_{0}^{2}-q^{2}}{\sqrt{1-x^{2}}}\delta(q_{0}^{2}
-2q_{0}E_{\textbf{k}}-q^{2}-2kqx)
\end{equation}
And in the above equations, we have defined
\begin{equation}
Re f(q_{0})=[f(q_{0})+f(-q_{0})]/2 ,Re
\delta(f(q_{0}))=[\delta(f(q_{0}))-\delta(f(-q_{0}))]/2 \nonumber
\end{equation}

If we set $m=0$ , we calculate explicitly and find they just
coincide with the Ref.\cite{wunsch2006NJP}\cite{Hwang2007PRB}, and
its finite temperature counterparts were discussed in
Ref.\cite{Vafek2007PRL}\cite{Asgari2007JPA} while we emphasize that
within our model we do not need the overlapping factor used in their
work and the results come out more naturally.
\begin{equation}
\Im m\Pi_{matter}^{ret}=\frac{1}{\pi}Re\int_{0}^{k_{F}}\ dk
\int_{-1}^{1}
dx\frac{4k^{2}-4kq_{0}+q_{0}^{2}-q^{2}}{\sqrt{1-x^{2}}}\delta(q_{0}^{2}
-2kq_{0}-q^{2}-2kqx)
\end{equation}

\begin{eqnarray}
&\Im m\Pi_{matter}^{ret}& =\frac{1}{\pi}\int_{0}^{k_{F}}\ dk
[\frac{(q_{0}-2k)^{2}-q^{2}}{q^{2}-q_{0}^{2}}]^{\frac{1}{2}}\{\Theta(q-q_{0})\Theta(k-\frac{q+q_{0}}{2})\nonumber\\
&&+\Theta(q_{0}-q)[\Theta(\frac{q_{0}+q}{2}-k)-\Theta(\frac{q_{0}-q}{2}-k)]\}\nonumber\\
&&-\frac{1}{\pi}\int_{0}^{k_{F}}\ dk
[\frac{(q_{0}+2k)^{2}-q^{2}}{q^{2}-q_{0}^{2}}]^{\frac{1}{2}}
\Theta(q-q_{0})\Theta(k-\frac{q-q_{0}}{2})
\end{eqnarray}
\begin{equation}
\Pi_{matter}^{ret}=\Theta(q_{0}-q)\Pi_{1}^{+}+\Theta(q-q_{0})\Pi_{2}^{+}
\end{equation}

\begin{eqnarray}
\Im m\Pi_{1}^{+}&=&-\frac{1}{2\pi\sqrt{q_{0}^{2}-q^{2}}}\{(2k_{F}-q_{0})\sqrt{q^{2}-(2k_{F}-q_{0})^{2}}+q^{2}\arcsin(\frac{2k_{F}-q_{0}}{2k_{_{F}}})
\Theta(q-|q_{0}-2k_{F}|)\nonumber\\
&+&\frac{\pi
q^{2}}{2}[\Theta(2k_{F}-q_{0}-q)+\Theta(2k_{F}-q_{0}+q)]\}
\end{eqnarray}

\begin{eqnarray}
\Im m\Pi_{2}^{+}&=&
\frac{\Theta(2k_{F}+q_{0}-q)}{2\pi\sqrt{q^{2}-q_{0}^{2}}}\{(2k_{F}+q_{0})\sqrt{(2k_{F}+q_{0})^{2}-q^{2}}-q^{2}\ln\frac{\sqrt{(2k_{F}+q_{0})^{2}-q^{2}}+2k_{F}+q_{0}}{q}\nonumber\\
&-&[(2k_{F}-q_{0})\sqrt{(2k_{F}-q_{0})^{2}-q^{2}}-q^{2}\ln\frac{\sqrt{(2k_{F}-q_{0})^{2}-q^{2}}+2k_{F}-q_{0}}{q}]\Theta(2k_{F}-q_{0}-q)\}\nonumber\\
\end{eqnarray}

The real part can be obtained using Kramers-Kronig relation
\begin{equation}
\Re e\Pi_{matter}^{ret}=\frac{1}{\pi}\int_{-\infty}^{\infty}
dq_{0}'\frac{\Im m\Pi_{matter}^{ret}(q_{0})}{q_{0}'-q_{0}}
\end{equation}
For the massive case(gapped graphene), the explicit calculation of
the polarization function is
straightforward\cite{asgari2009prb}\cite{Pyatkovskiy2009jpcm}, so we
just discuss an important static case $q_{0}=0$

\begin{eqnarray}
\Pi_{matter}^{ret}(\textbf{q},0)&=&\frac{2}{\pi^{2}}\int_{0}^{k_{F}}\frac{d
k
k}{E_{\textbf{k}}}\int_{-1}^{1}dx\frac{1}{\sqrt{1-x^{2}}}(\frac{q^{2}-4E_{\textbf{k}}^{2}}
{q^{2}+2kq x}-1)\nonumber\\
&=&-\frac{2}{\pi}(\sqrt{k_{F}^{2}+m^{2}}-\sqrt{m^{2}})\nonumber\\
&&+\frac{2}{\pi q}\int_{0}^{k_{F}}\frac{dk
k}{\sqrt{k^{2}+m^{2}}}(\sqrt{q^{2}-4k^{2}}-4m^{2}/\sqrt{q^{2}-4k^{2}})\Theta(\frac{q}{2}-k)\nonumber\\
&=&-\frac{2}{\pi}(\mu-m)\nonumber\\
&&+\frac{2}{\pi
q}\int_{m}^{\mu}d\epsilon(\sqrt{q^{2}+4m^{2}-4\epsilon^{2}}\nonumber\\
&&-4m^{2}/\sqrt{q^{2}+4m^{2}-4\epsilon^{2}})\Theta(\sqrt{q^{2}+4m^{2}}-2\epsilon)
\end{eqnarray}

\begin{eqnarray}
&&\frac{2}{\pi
q}\int_{m}^{\mu}d\epsilon(\sqrt{q^{2}+4m^{2}-4\epsilon^{2}}-4m^{2}/\sqrt{q^{2}+4m^{2}-4\epsilon^{2}})\Theta(\sqrt{q^{2}+4m^{2}}-2\epsilon)\nonumber\\
=&&\frac{1}{\pi
q}\{[-mq+\frac{q^{2}-4m^{2}}{2}\arctan\frac{q}{2m}]\Theta(2k_{F}-q)\nonumber\\
&&+[\mu\sqrt{q^{2}-4k_{F}^{2}}-mq-\frac{q^{2}-4m^{2}}{2}\arctan\frac{\sqrt{q^{2}-4k_{F}^{2}}}{2\mu}+\frac{q^{2}-4m^{2}}{2}\arctan\frac{q}{2m}]\Theta(q-2k_{F})\}\nonumber\\
\end{eqnarray}

\section{Static Screening}

The static screening properties of the massive Dirac fermions in the
RPA are controlled by the static dielectric function
$\varepsilon_{RPA}(q,0)$.
\begin{equation}
\varepsilon_{RPA}(q,0)=1-\frac{2\pi e^{2}}{\kappa
q}[\Pi_{vac}(\textbf{q},0)+\Pi_{matter}(\textbf{q},0)]
\end{equation}
Without the matter contribution, the polarized charge distribution
has been discussed in Ref.\cite{Kotov2008PRB} ,if we take into account the matter part, things will be quite different

For $q\leq2k_{F}$\\
\begin{equation}
\Pi(\textbf{q},0)=-\frac{2\mu}{\pi}
\end{equation}

For$q>2k_{F}$\\
\begin{eqnarray}
\Pi(\textbf{q},0)=-\frac{2\mu}{\pi}+\frac{1}{\pi
q}\Theta(q-2k_{F})[\mu\sqrt{q^{2}-4k_{F}^{2}}-\frac{q^{2}-4m^{2}}{2}\arctan\frac{\sqrt{q^{2}-4k_{F}^{2}}}{2\mu}]
\end{eqnarray}
Similar results has been obtained in Ref.\cite{Gorbar2002PRB}.  From
above we can deduce that the density of state at the Fermi surface
is given by $D(k_{F})=2\mu/\pi$, In the two limiting cases, the
total static polarizability becomes a constant as in normal 2D
electron liquid systems and 2D massless Dirac fermion systems. For
2D electron liquid, the 2D Thomas-Fermi wave vector is given by
$q_{TF}=me^{2}/\kappa$. Consider a external charge density embedded
in the homogeneous fermions.Even though
$V_{ei}(\textbf{x})=C\delta(\textbf{x})$ is short-ranged, the
density response is not short-ranged which can be calculated by
linear response theory\cite{Giuliani2005QTEL}.
\begin{equation}
\Pi(\textbf{x})=C\int\frac{d^{2}\textbf{q}}{(2\pi)^{2}}\Pi(\textbf{q},0)e^{i\textbf{q}\cdot\textbf{x}}
\end{equation}
It is interesting to note that as the normal 2D electron liquid, the
2D Dirac fermions also has an oscillatory term due to the
non-analyticity at $q=k_{F}$. Next, we take into account finite
temperature dependant screening, the polarization function at
$T\neq0$ is
\begin{eqnarray}
&&\Pi_{matter}^{ret}(\textbf{q},T)=-\frac{2}{\pi}\{\mu-m+\frac{1}{\beta}\ln[1+e^{-\beta(\mu-m)}]+\frac{1}{\beta}\ln[1+e^{-\beta(\mu+m)}]\}\nonumber\\
&&+\frac{2}{\pi
q}\int_{m}^{\epsilon_{q/2}}d\epsilon(2\sqrt{\epsilon_{q/2}^{2}-\epsilon^{2}}-2m^{2}/\sqrt{\epsilon_{q/2}^{2}-\epsilon^{2}})[\frac{1}{1+e^{\beta(\epsilon-\mu)}}+\frac{1}{1+e^{\beta(\epsilon+\mu)}}]\nonumber\\
\end{eqnarray}
At low temperature $(T\ll T_{F})$,we have the asymptotic from of the
polarizability from the above equation
\begin{eqnarray}
\Pi(\textbf{q},T)\approx-\frac{2\mu(T)}{\pi}  &\textrm{for} &
\epsilon_{q/2}<\mu,
\end{eqnarray}
In particular, for $q=2k_{F}$, $m\rightarrow0$, we have
\begin{eqnarray}
\Pi_{matter}^{ret}(\textbf{q},T)=&&-\frac{2}{\pi}\{\mu+\frac{1}{\beta}\ln[1+e^{-\beta\mu}]+\frac{1}{\beta}\ln[1+e^{-\beta\mu}]\}\nonumber\\
&&+\frac{4}{\pi
q}\int_{0}^{\epsilon_{q/2}}d\epsilon \sqrt{\epsilon_{q/2}^{2}-\epsilon^{2}} [\frac{1}{1+e^{\beta(\epsilon-\mu)}}+\frac{1}{1+e^{\beta(\epsilon+\mu)}}]\nonumber\\
\approx
&&-\frac{2}{\pi}\mu(T)+\frac{2k_{F}}{\pi}\int_{0}^{1}dx\sqrt{1-x^{2}}(1-e^{\beta(k_{F}x-\mu)}+e^{2\beta(k_{F}x-\mu)}+...)\nonumber\\
\approx &&-\frac{2}{\pi}\mu(T)+\frac{k_{F}}{2}-\sqrt{\frac{2}{\pi
k_{F}}}(1-\frac{\sqrt{2}}{2})\zeta(\frac{3}{2})T^{3/2}
\end{eqnarray}
And the total polarization function is
\begin{eqnarray}
\Pi(\textbf{q},T)\approx -\frac{2}{\pi}\mu(T)-\sqrt{\frac{2}{\pi
k_{F}}}(1-\frac{\sqrt{2}}{2})\zeta(\frac{3}{2})T^{3/2}
\end{eqnarray}
While for a general $m$, analytic expressions can not be obtained,
however, it is quite easy to obtain numerical results from our
semi-analytical results.

At high temperature $(T\gg T_{F})$
\begin{eqnarray}
\Pi(\textbf{q},T)\approx \frac{2}{\pi}\mu(T)+\frac{1}{\pi
T}[\frac{q^{2}}{12}-m^{2}]
\end{eqnarray}
For the chemical potential $\mu(T)$, we can get its explicit form
the conservation of the total electron density. For the massless
Dirac fermions, the asymptotic expression has been given in
Ref.\cite{Hwang2009PRB}. We note that for the massive case the
expressions is the same.

\section{Conclusion}
In summary, we have presented a finite temperature field model for
the 2D massive Dirac fermions and calculated the polarization
functions for massive Dirac fermions. Finite temperature and finite
density were taken into account to analyze the physics of gapped
graphene in a more general case and connect various other group's
important
work\cite{wunsch2006NJP}\cite{Hwang2007PRB}\cite{Vafek2007PRL}\cite{Asgari2007JPA}
done before. These results may be useful to study finite-temperature
screening within the RPA. Other important quantities, for example
conductivity or specific heat, can be obtain from our results.

The polarization function at finite temperature can be also used to
calculate the thermodynamic properties of massive Dirac fermion.\\

 \end{document}